\documentclass{aa}
\usepackage{graphics}
\def\deg{$^{\circ}\,$}

\begin{document}

\thesaurus{   (11.05.2;  
               11.06.1;  
               11.09.2;  
               11.11.1:  
               11.19.4)}

\authorrunning{Conselice, Bershady \& Gallagher}

\title{Physical morphology and triggers of starburst galaxies}

\titlerunning{Physical morphology of starburst galaxies}

\author{C.J. Conselice, M.A. Bershady, J.S. Gallagher, III}

\offprints{C.J. Conselice}

\institute{Department of Astronomy, The University of Wisconsin-Madison \\
Madison WI., 53706}
\date{Accepted to A\&A Letters}
\maketitle

\begin{abstract}

We present a method of determining the likely triggering mechanisms
for luminous, nearby starbursts based on their optical asymmetry and
(B-V) color -- what we term as indices of `physical morphology.' By
comparing the starbursts' locations in a color-asymmetry diagram to an
independent, nearby galaxy sample, we identify which starbursts are
likely triggered by an interaction/merger. We confirm our
morphological interpretation based on a comparison to 20\% and 50\% HI
velocity-width ratios. We further explain how this use of physical
morphology can put constraints on the dynamical history of other
galaxies, including star-forming galaxies at high redshifts.

\keywords{starburst galaxies -- high-redshift galaxies --
galaxy parameters - color, asymmetry, line widths.}
\end{abstract}

\section{Introduction}

The large number of galaxies at high redshifts ($z$) undergoing
intense star-formation (Steidel et al. 1996, Lowenthal et al. 1997)
suggest starburst galaxies were a dominant phase of early galaxy
evolution. While the rate and specific intensity of star formation in
these distant galaxies is certainly higher than typical, nearby
starbursts (Weedman et al. 1999), local starbursts are similar to
star-forming high-$z$ galaxies in terms of their structural and
stellar characteristics (Giavalisco et al. 1996; Hibbard \& Vacca
1997; Heckman et al. 1998; Conselice et al. 2000a).

Related issues include determining how starbursts are triggered, and
if the triggering mechanisms change with $z$. Starburst triggering
mechanisms include: interactions and mergers (Schweizer 1987; Jog \&
Das 1992), bar instabilities (Shlosman et al. 1990), and kinematic
effects from SNe and stellar winds (e.g.  Heckman et al. 1990). While
the merging is expected to be more common at earlier epochs, the first
epoch of star-formation could occur as a result of the inital collapse
of individual gas clouds.

For nearby, luminous galaxies, usually it is possible to determine
what triggers a starburst by examining kinematic, and pan-chromatic
structural information.  This can be quite expensive in telescope
time, particularly for high-$z$ starbursts, where detailed
spectroscopic information is difficult to obtain at present.  While
some starbursts are undergoing interactions or mergers, it is
difficult to quantify the strength and youth of such events. A method
of determining starburst triggers based on morphology or other, easily
observable properties of a galaxy would be ideal. In this paper we
present a method to determine objectively if a starburst is triggered
by a galaxy interaction based on its color and $R$-band asymmetry.
 
\section{The Sample and Optical Data}

The sample consists of five UV-bright examples of nearby galaxies
chosen from a study of northern hemisphere starbursts: Markarian 8,
NGC 3310, NGC 3690, NGC 7673, NGC 7678.  These galaxies were imaged in
several bands with the WIYN 3.5m telescope\footnote{The WIYN
Observatory is a joint facility of the University of
Wisconsin-Madison, Indiana University, Yale University, and the
National Optical Astronomy Observatories.}, located at the Kitt Peak
National Observatory, using a 2048$^{2}$-pixel thinned SB2K CCD with a
$6.8\times 6.8$ arcmin$^{2}$ field of view and a scale of 0.2 arcsec per
pixel. The seeing during the observations on average was $1''$ FWHM.
The $R$-band images used here are bias subtracted, flat-fielded, and
cleared of foreground stars and background galaxies.  These
contaminating objects typically can cause rather high asymmetries if
not properly removed.  The (B-V) colors for all galaxies are from the
RC3 catalog, except NGC 3690, where we adopt Weedman's (1973) value.

Each of the starbursts in our sample are benchmarks; i.e. they 
are relatively well studied and understood.
 
\begin{figure} 
\resizebox{\hsize}{!}{\includegraphics{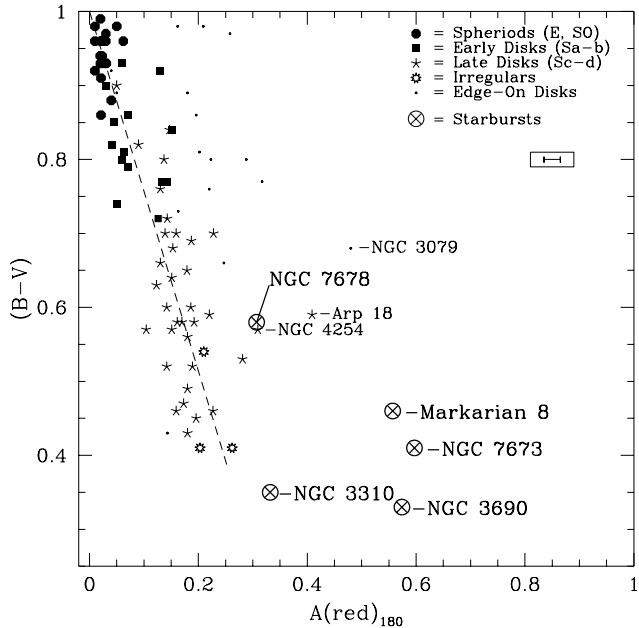}}
\caption{Color-Asymmetry diagram for the Frei et al. (1996) and
starburst samples. Extreme outliers and starbursts are labeled.}
\end{figure}

{\bf Markarian 8} hosts an intense starburst, with very blue colors
(Huchra 1977). This galaxy contains several distinct `pieces' with
visible tidal tails, and has long been recognized as a merger, or a
strongly interacting double galaxy (Casini \& Heidmann 1976; Keel \&
van Soest 1992). The interaction/merger between the multiple
components of this galaxy are responsible for triggering the
star-formation in the disk.

{\bf NGC 3310}, classified as a barred spiral (RC3), has it's very
young starburst in a 1 kpc diameter ring around the nucleus.  The bar
and the size of the ringed structure suggest this starburst was
triggered by a bar instability (Athanassoula 1992; Piner et
al. 1995). Faint outer ripples are evidence for a minor merger or
interaction with another galaxy, probably a dwarf (e.g. Balick \&
Heckman 1981; Schweizer \& Seitzer 1988); this plausibly produced the
bar instability which led to the starburst.

{\bf NGC 3690}, along with Markarian 8 are the most extreme
interactions/mergers in our sample.  The second `half' of NGC 3690 is
IC 694 (e.g. Gehrz et al. 1983).  There is no disk structure to this
galaxy, which is populated by very luminous, high surface-brightness
star forming regions (Soifer et al. 1989).

{\bf NGC 7673} morphologically consists of an inner disturbed spiral
structure, faint outer ripples (Homeier \& Gallagher 1999), and huge,
blue star-forming clumps embedded in a disturbed H I disk (Nordgren et
al. 1997). These features are clues that this galaxy recently
interacted with another galaxy, triggering the starburst.

{\bf NGC 7678}, classified as a barred spiral, contains a starburst
located in a roughly symmetrical spiral pattern, similar to NGC 3310.
This starburst consists of several bright H II regions (Goncalves et
al. 1998) and contains a Seyfert nucleus (Kazarian 1993).  NGC 7678
contains a large, massive blue arm where much of the starburst is
located (Vorontsov-Velyaminov 1977).

Most of these galaxies show evidence for an interaction, but in
various degrees and intensities. NGC 3310, NGC 7678 are
probably minor mergers, or interactions that occurred in the distant
past, while NGC 3690, NGC 7673, and Markarian 8 are obvious collisions that
contain very disturbed structures.


\section{Asymmetry}

The asymmetry method used here, described in detail by Conselice,
Bershady \& Jangren (2000), gives a simple, quantified measure of how
a galaxy deviates from perfect axi-symmetry.  Like Abraham et
al. (1996), our algorithm consists of rotating a galaxy 180\deg about
a center, subtracting the rotated image from the original, and
dividing the sum of the absolute value of the pixels in the residual
image by the sum of pixel values in the original image. The higher
the intensity of the residuals, the larger the asymmetry. In our
implementation, however, (a) we repeat the asymmetry computation using
different center estimates until a minimum asymmetry is found, and (b)
we make a correction for noise.

The rotational asymmetry of normal galaxies increases with the
`lateness' of their morphological type (Conselice, 1997), indicating
that at least some component of their asymmetry is associated with the
flocculent appearance of star-formation within galaxies. Asymmetry may
also arise, however, from large-scale, dynamical perturbations. Other
methods of asymmetry measurement (e.g. Zaritsky \& Rix, 1997;
Kornreich, Haynes, \& Lovelace, 1998) are particularly sensitive to
this dynamical component. In contrast, our rotational measurement is
sensitive to both dynamical and flocculent components of asymmetry. As
we show in the next section, interacting and irregular galaxies can be
distinguished on the basis of the relative amplitudes of these two
asymmetry components via the color-asymmetry diagram. Asymmetry is a
powerful quantitative morphological parameter, and in conjunction with
the color of a galaxy, it can be used to determine the physical nature
of a galaxy.

\begin{figure}
\resizebox{\hsize}{!}{\includegraphics{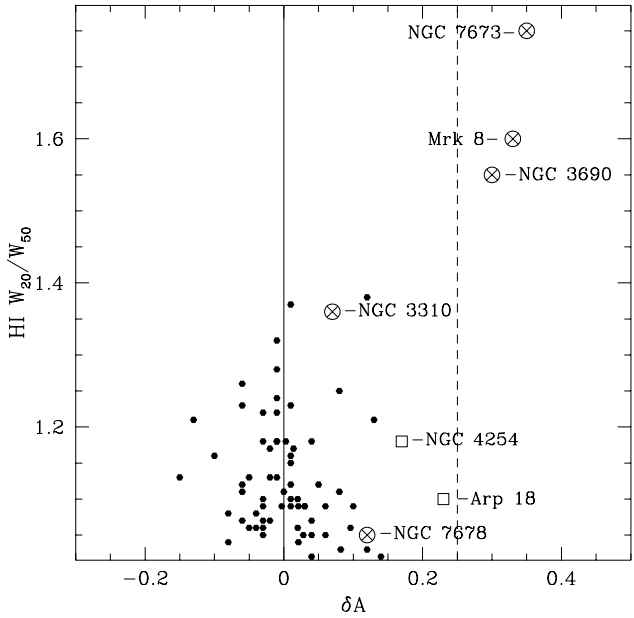}}
\caption{Ratio of HI line-widths at 20\% and 50\% levels vs.
asymmetry-offset from fiducial sequence of Figure 1 (Frei et al. sample -
dots and open squares; starbursts - $\otimes$).}
\end{figure}

\section{The Color-Asymmetry Diagram (CAD)}

The CAD for a sample of 113 nearby galaxies (Frei et al. 1996;
Conselice et al. 2000b) and the starburst sample is shown in Figure
1. A correlation between asymmetry and color can be seen for most
Hubble types: Early type galaxies (E, S0) populate the symmetric, red
corner of the diagram; later type galaxies become progressively bluer
{\it and} more asymmetric. This `fiducial' trend with Hubble type is
indicated by the dashed line in Figure 1. Some objects do not lie on
the fiducial color-asymmetry sequence. Most of these are
highly-inclined systems, as discussed by Conselice et al. 2000b.
However, visual inspection of the most extreme outliers at blue colors
of B-V$<$0.7 (i.e., NGC 3079, Arp 18, and NGC 4254) indicates they are
are too asymmetric for their colors because they are dynamically
disturbed. It appears to be possible to distinguish if a galaxy is
interacting or merely undergoing normal star formation (such as
irregular galaxies) based simply on its position in the CAD.
 
Where do the starbursts lie in the CAD? Their CAD positions are mostly
consistent with a strong interaction, or merger origin (Figure 1,
$\otimes$). Moreover, the amount of deviation from the fiducial galaxy
sequence appears to be correlated roughly with the degree of
interaction/merger. For example, NGC 3690 and Markarian 8 are major
mergers between galaxies of similar sizes, and have two of the most
deviant positions.  NGC 3310, on the other hand, almost fits along the
fiducial galaxy sequence; this starburst is believed to be produced by
a bar instability or minor interaction (Conselice et al. 2000a). The
fact that NGC 3310 does not fit exactly along this sequence is
probably due to the fact that it was involved with a minor merger in
the past (Balik \& Heckman 1981).

We can quantify and test our interpretation of the CAD by comparing a
starbursts deviance in the CAD with a kinematic indicator of
deviance. A variety of HI studies show that interacting galaxies often
have unusual and asymmetric global HI emission line profiles, often
with extended velocity tails (e.g., Gallagher, Knapp, \& Faber 1981,
Bushouse 1987, Mirabel \& Sanders 1988, Sancisi 1997). Here we introduce the 
use of the ratio of HI line-width at 20\% and 50\% maximum, as extracted 
from the Lyon-Meudon Extragalactic Database (LEDA) as a new dynamical 
indicator. High values of W$_{20}$ / W$_{50}$ imply shallower rising HI 
profiles or wings, and hence should be an indicator of a recent dynamical
disturbance.

As Figure 2 shows, this line-width ratio is large for the starburst
galaxies with asymmetries that deviate from the fiducial sequence.
Galaxies shown by the color-asymmetry digram to be strongly
interacting or merging, namely NGC 7673, NGC 3690 and Markarian 8, all
have the largest HI line width ratios. NGC 3310 and NGC 7678, which
have symmetric inner spiral structures and are probably older
starbursts perhaps triggered by bar instabilities, have smaller
line-width ratios and are less deviant from the fiducial sequence in
color and asymmetry. Hence our interpretation of starburst origin
based on the color-asymmetry diagram is corroborated.

Finally, we note that the starburst galaxies with high asymmetries
have blue UBV colors (Figure 3), but do not lie outside the range for
normal galaxies (rectangle, Figure 3, adopted from Huchra, 1977).
Larson and Tinsley (1978) showed that galaxies with tidal features,
such as tidal tails, had a large scatter on a UBV color-color
plots. This is not evident for our small sample, which indicates that
the color-asymmetry method is more sensitive to identifying
merger-induced starbursts than colors alone.

\section{Implications for High-Redshift Galaxies}

The origins and triggering mechanisms of the increasingly abundant and
luminous starbursts observed at intermediate and high $z$ are
still an issue of debate. Merging is invoked as a likely candidate for
triggering because the physical volume of the universe decreases at
earlier times, however this has not been demonstrated directly. The
physical-morphological method outlined in this paper can be used to
determine if merging is indeed the culprit. If these star-forming
galaxies at high $z$ are triggered by interactions/mergers then
their positions in the CAD would be similar those for the starbursts
presented in this paper. If the interactions/mergers are minor, then
their locations would fall near the normal galaxy fiducial sequence,
similar to NGC 3310.

\begin{figure}
\resizebox{\hsize}{!}{\includegraphics{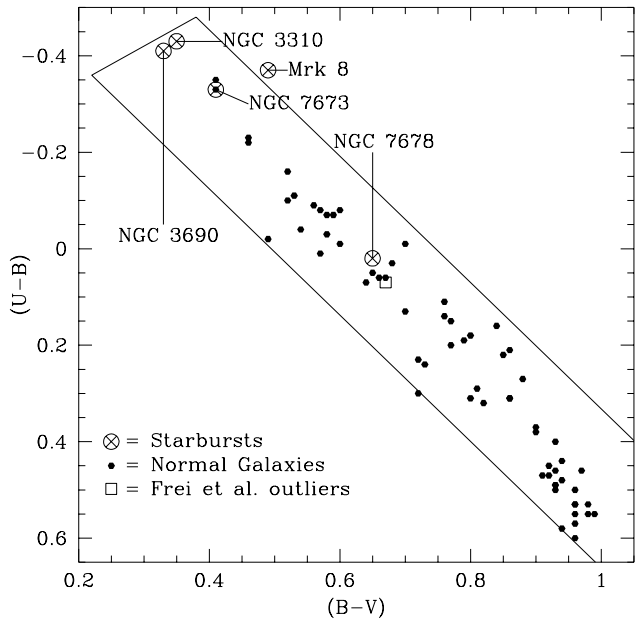}}
\caption{UBV diagram for Frei et al. and starburst samples.}
\end{figure}

In principle, the high resolution of the Hubble Space Telescope allows
detailed morphological studies of distant galaxies. For example,
asymmetry has previously been used in conjunction with the
concentration of light for galaxies in the Hubble Deep Field (Abraham
et al. 1996) and other WFPC-2 images (Jangren et al. 1999).  The
method used to compute the asymmetries in Abraham et al.'s study
differ in several important respects from the method used here and by
Jangren et al. In particular, Conselice et al. (2000b) demonstrated
that high angular resolution ($<$ 0.1 arcsec FWHM) is critical for the
study of distant galaxy asymmetry. While deep NICMOS images of the
Hubble Deep Field allow an unprecedented opportunity to study the rest
frame optical morphology of high-$z$ galaxies, adaptive optics on
large, ground-based telescope or the Next Generation Space Telescope
may ultimately prove critical for the study of asymmetries in distant
galaxies.

\section{Conclusions}

In this letter we have demonstrated a morphological method of
deciphering the triggering mechanisms of starbursts galaxies through
the use of a color-asymmetry diagram. Based on a sample of nearby
starbursts, we have demonstrated that those starbursts generally
regarded as triggered by interactions or mergers are located in a
special region of CAD characterized by large optical asymmetry at a
given color. In contrast, the two starbursts in our sample (NGC 3310
and NGC 7678) that are probably not triggered from a major
interaction/merger fall nearly on the fiducial color-asymmetry
sequence of normal galaxies.

While our sample is small, we suggest that {\it the color-asymmetry
diagram can be used to separate starbursts triggered by
mergers/interactions from those triggered from other causes.} We
confirm this interpretation by comparing the degree of deviation from
the fiducial color-asymmetry sequence to HI line-width ratios, which
serves as a dynamical indicator of strong gravitational interactions.
This physical-morphological method is quantitative and appears to be
superior to analyses of outliers in, e.g., two-color plots such as
$U-V$ and $B-V$. This physical-morphological method is also
well-suited for studying the nature and evolution of distant galaxies
where it is difficult to gather information beyond images, magnitudes,
and redshifts.

This work was funded by NASA contract WAS7-1260 to JPL; AR7539, AR7518
and GO7875, and GO7339 from STScI which is operated by AURA,
Inc. under NASA contract NAS5-26555; NASA LTSA contract NAG5-6032; and
NSF contract AST-9970780. JSG and CJC thank the WFPC-2 team.

\end{document}